\documentclass[ reprint, superscriptaddress, amsmath,amssymb,
aps, pra, showkeys ]{revtex4-1}

\usepackage[english]{babel}

\usepackage[utf8]{inputenc}
\usepackage{booktabs}   
\usepackage{array}      

\AtBeginDocument{
    
    \newwrite\bibnotes
    \def\bibnotesext{Notes.bib}
    \immediate\openout\bibnotes=\jobname\bibnotesext
    \immediate\write\bibnotes{@CONTROL{REVTEX41Control}}
    \immediate\write\bibnotes{@CONTROL{apsrev41Control,author="48",editor="1",pages="0",title="0",year="1"}}

     \if@filesw
     \immediate\write\@auxout{\string\citation{apsrev41Control}}%
    \fi

}

\usepackage{natbib}
\usepackage{caption}
\usepackage[utf8]{inputenc}
\usepackage{amsthm}
\usepackage{mathtools}

\usepackage{xcolor}
\usepackage{graphicx}
\usepackage[left=23mm,right=13mm,top=35mm,columnsep=15pt]{geometry} 
\usepackage{adjustbox}
\usepackage{placeins}
\usepackage[T1]{fontenc}
\usepackage{lipsum}
\usepackage{csquotes}
\usepackage{comment}
\usepackage[normalem]{ulem}
\usepackage{float}
\usepackage{siunitx} 
\usepackage{multirow}
\usepackage{booktabs}
\usepackage{siunitx}
\usepackage{array}
\usepackage{graphicx} 
\usepackage{colortbl}
\usepackage{pifont}
\usepackage{array}

\usepackage{booktabs} 

\definecolor{SoftRed}{RGB}{230,80,80} 
\definecolor{SoftOrange}{RGB}{245,170,80} 
\definecolor{SoftGreen}{RGB}{100,200,100} 

\usepackage[pdftex, pdftitle={Article}, pdfauthor={Author}]{hyperref} 
\hypersetup{
    colorlinks,
    linkcolor={blue!50!blue},
    citecolor={blue!50!blue},
    urlcolor={blue!80!black}
}

\makeatletter
\@ifclassloaded{revtex4-1}{\typeout{RevTeX 4.1 is loaded}}{\typeout{RevTeX 4.1 is NOT loaded}}
\makeatother

\begin{document}
\title{All-dry pick-up and transfer method for quantum emitter arrays in hexagonal boron nitride}

\author{Mohammad N. Mishuk}
    \email[Electronic mail: ]{nasim.mishuk@tum.de}
\affiliation{Department of Computer Engineering, TUM School of Computation, Information and Technology, Technical University of Munich, 80333 Munich, Germany}
\affiliation{Munich Center for Quantum Science and Technology (MCQST), 80799 Munich, Germany}
\affiliation{Abbe Center of Photonics, Institute of Applied Physics, Friedrich Schiller University Jena, 07745 Jena, Germany}

\author{Mouli Hazra}
\affiliation{Department of Computer Engineering, TUM School of Computation, Information and Technology, Technical University of Munich, 80333 Munich, Germany}
\affiliation{Munich Center for Quantum Science and Technology (MCQST), 80799 Munich, Germany}
\affiliation{Abbe Center of Photonics, Institute of Applied Physics, Friedrich Schiller University Jena, 07745 Jena, Germany}
\author{Anand Kumar}
\affiliation{Department of Computer Engineering, TUM School of Computation, Information and Technology, Technical University of Munich, 80333 Munich, Germany}
\affiliation{Munich Center for Quantum Science and Technology (MCQST), 80799 Munich, Germany}
\affiliation{Abbe Center of Photonics, Institute of Applied Physics, Friedrich Schiller University Jena, 07745 Jena, Germany}
\author{Peter Dannberg}
    \affiliation{Fraunhofer Institute for Applied Optics and Precision Engineering (IOF), 07745 Jena, Germany}      
\author{Aslı Çakan}
\affiliation{Department of Computer Engineering, TUM School of Computation, Information and Technology, Technical University of Munich, 80333 Munich, Germany}
\affiliation{Munich Center for Quantum Science and Technology (MCQST), 80799 Munich, Germany}

\author{Tobias Vogl}
    \email[Author to whom correspondence should be addressed: ]{tobias.vogl@tum.de}
\affiliation{Department of Computer Engineering, TUM School of Computation, Information and Technology, Technical University of Munich, 80333 Munich, Germany}
\affiliation{Munich Center for Quantum Science and Technology (MCQST), 80799 Munich, Germany}
\affiliation{Abbe Center of Photonics, Institute of Applied Physics, Friedrich Schiller University Jena, 07745 Jena, Germany}


\begin{abstract}
Single photon emitters in hexagonal boron nitride are based on fluorescent point-like defects. These defects typically have exceptional photophysical properties and therefore been the focus of extensive research due to their potential to advance photonic quantum technologies. However, achieving scalable integration of these emitters to arbitrary platforms with high yield while retaining their characteristics remains a significant challenge, particularly when the target substrate is not compatible with the fabrication method. In this work, we introduce an all-dry transfer method aimed at addressing these challenges with improved effectiveness compared to existing techniques. This polymer stamp-assisted transfer method maintains high output and preserves the fundamental characteristics of the emitters while eliminating wet chemical processes. A comprehensive post-transfer characterization verified not only the maintenance of the defining characteristic of a single photon emitter, the second-order correlation function $g^{(2)}(0)$, but also showed improvement by about 46\%. In contrast, the lifetime, emission spectrum, and the photostability showed only negligible change, demonstrating that the characteristics of the emitters were retained during the transfer process. This transfer technique has success rate of 81.8\%, determined by the proportion of single photon emitters that retain their optical and preserve physical structure post-transfer. This high success rate shows the potential to scale the integration of single photon emitters across diverse platforms. We expect that this process contributes to the applications of boron nitride defects in quantum technologies.
\end{abstract}


\maketitle

\section{Introduction} \label{sec:introduction}
Since the initial observation of single photon emission from the defects in monolayers and multilayers hexagonal boron nitride (hBN) in 2015 at room temperature \cite{tran_quantum_2016}, single photon emitters (SPEs) in hBN are widely researched within the field of study \cite{Cholsuk2024a}. These SPEs offer several advantages, including exceptional photostability \cite{li_prolonged_2023}, tunability and high-purity emission \cite{grosso_tunable_2017}, robustness \cite{10.1038/s41467-019-09219-5}, and high brightness \cite{gan_large-scale_2022} at room temperature \cite{tran_quantum_2016}. With these characteristics, SPEs in hBN are promising candidates in advancing quantum technologies such as quantum sensing, quantum cryptography, and quantum computing \cite{boretti_advancing_2025, liu_2d_2019,https://doi.org/10.1002/adom.202402508}. On demand creation and engineering of the SPEs are crucial for robust communication in quantum cryptography \cite{gisin_quantum_2002}, enhancing statistical precision in quantum metrology \cite{giovannetti_advances_2011}, and implementing the entangling logic gates to achieve the universal logic gates for quantum computing \cite{obrien_optical_2007}.

Over the last decade, the reliable and reproducible process of creating the SPEs in hBN has gained significant advancement \cite{tran_quantum_2016, bourrellier_bright_2016, bourrellier_bright_2016, mendelson_identifying_2021,gu_engineering_2021, xu_single_2018, vogl_fabrication_2018, liu_single_2024, kumar_localized_2023}. The process of creation of emitters mainly involves introducing defects in the crystal lattices of hBN by ion implantation \cite{mendelson_identifying_2021}, ion/electron irradiation \cite{kumar_localized_2023, gu_engineering_2021}, oxygen \cite{vogl_fabrication_2018} or argon \cite{xu_single_2018} plasma treatment, thermal annealing \cite{liu_single_2024} among other techniques. Recent studies have demonstrated the reproducible generation of bright and stable SPEs using electron beam irradiation methods \cite{kumar_localized_2023, choi2016engineering, ngoc2018effects}.

Despite their promising characteristics and various methods for defect engineering in hBN, the full potential of SPEs in optical quantum technologies remains underutilized due to challenges in their integration into diverse devices \cite{sartison_scalable_2022}. A transfer technique can be one of the promising approaches to overcome these challenges by precisely positioning pre-fabricated SPEs in different platforms and devices \cite{proscia2018near}. Enabling the integration of the SPEs in photonic structures like, waveguides and cavities, these transfer methods can potentially increase the functionality of the SPEs \cite{froch2022purcell, elshaari2020hybrid}. Various transfer techniques have been reported to tackle these challenges, each with its own advantages and limitations.

The integration techniques can be categorized into three ways \cite{sartison_scalable_2022}: They are monolithic integration, heterogeneous integration, and hybrid integration. In the monolithic integration technique, the devices are made of the same material that hosts the quantum emitters \cite{li2021integration}; however, the devices are limited by performance issues induced by the intrinsic material properties. For example, the device can perform below the optimum level due to low electrical conductivity and optical gain \cite{sartison_scalable_2022, kaur_hybrid_2021}. On the other hand, in the heterogeneous technique wafer bonding is required to bond different materials. This technique allows the integration of light sources like quantum dots into silicon-based photonic integrated circuits (PICs) \cite{davanco2017heterogeneous}. However, the technique is limited by high bonding temperature (over 300°C) which can cause thermal stress impacting the versatility of the integration process for temperature sensitive devices \cite{sartison_scalable_2022, ke_review_2020, plosl_wafer_1999}.

In this paper, we focus on hybrid integration technique, as it offers the independence of optimizing and combining different materials and platforms \cite{kim_hybrid_2020}. Hybrid integration enables the efficient creation of emitters on one platform and subsequent transfer of them to another platform. This flexibility paves an adaptable approach for advancing the hBN hosted SPE integration in the quantum domain. Over the years techniques such as wet transfer, dry transfer, semi-dry transfer, direct growth, inorganic membrane-assisted transfer, adhesive layer transfer etc. hybrid integration techniques were reported \cite{10.1088/1361-6463/aa7839,Ahmadi2024, proscia2018near, lu2016universal, scheuer2021polymer, kim2020review, nakatani2024ready, pelgrin2023hybrid, pham2024transfer}. Among the first demonstrated techniques was a wet deterministic transfer process of hBN defect-based quantum emitters created by oxygen plasma etching \cite{vogl_fabrication_2018}. The disadvantage is that polymers and solvents are used that typically induce contamination which can be optically active, therefore reducing the optical quality of the emitters \cite{vogl_fabrication_2018}. Besides residues, the introduction of new defects \cite{doi:10.1021/acsaom.4c00200}, trapped bubbles, water, and weak Van der Waals interactions between the material and the substrate are reported in many articles \cite{suk_transfer_2011, lin_graphene_2012, pirkle_effect_2011}.  For example, the supporting layers like polymethyl methacrylate can leave their residues on the hBN flake after the transfer, which might worsen the optical characteristics of the SPEs \cite{frisenda_recent_2018}. Moreover, trapped bubbles and water can form scattering centers and change the regional refractive index which can deter the performance of device \cite{dean_boron_2010}.

After reviewing the established processes of creation and transfer of single photon emitters we found out two major challenges that should be addressed to effectively integrate the SPEs: (i) It is challenging to create SPEs on a delicate device depending on the fabrication technique. To fabricate deterministic arrays of quantum emitters, localized irradiation with charged particles has been found to be effective \cite{kumar_localized_2023, fournier2021position, gale2022site}, however, these techniques typically only work on conducting substrates to avoid surface charging (that deflects the particle beam). In addition, high energy particles can damage the components like waveguides which results in the altercation of the optical properties of these devices \cite{exarhos_optical_2017}. On the other hand, ion implantation or laser irradiation etc. processes inadvertently bring out unintended stress and defects \cite{sontheimer_photodynamics_2017}. Consequently, a non-destructive and uncomplicated approach is instrumental for the integration of SPEs on sensitive devices. (ii) Current transfer techniques are complex and involve multiple chemical processing steps, such as wet etching and solvent cleaning, which can introduce contaminants and are difficult to scale for mass production \cite{castellanos-gomez_deterministic_2014}. These chemical methods can also affect surface chemistry and introduce variability in the SPEs' performance, hindering the reproducibility required for practical applications \cite{frisenda_recent_2018}. Simplifying the transfer process by minimizing chemical involvement is crucial for developing a scalable and reliable integration method.

A recent study by O'Hara et al.~\cite{o2024transfer} introduced a method for transferring pre-activated hBN emitters onto arbitrary substrate without any temperature offering valuable insight for the scalable integration hBN emitters. This insight aligns with our independent approach to take it to another step ahead to the seamless integration of hBN emitters into quantum photonic devices.
 
Here, we outline a strategy to help overcome the integration challenge of hBN-based SPEs. An all-dry temporary transfer stamp (TTS) was utilized which facilitates deterministic transfer of the hBN flake with high yield - a desired aspect for the integration of SPEs in PICs. Through careful characterization before and after the transfer process, we demonstrate that our technique maintains the photostability and emission purity of the SPEs, making it a viable solution for large-scale integration into quantum systems.  

\section{Experimental Methods} \label{sec:methods}
To test our transfer method, we first created arrays of SPEs on thin hBN flakes and investigated their optical and physical properties. Emitter arrays are necessary to study sufficient statistics of transferred emitters to be able to judge the success probability of different process conditions.

\subsection{Sample fabrication}
The hBN crystals (obtained from HQ Graphene) were mechanically exfoliated using the scotch tape method onto a viscoelastic polydimethylsiloxane (PDMS) polymer film (obtained from GelPak) to produce thin flakes. These flakes were then examined using optical microscopy and phase shifting interferometry (PSI) to identify flakes with a suitable thickness in the range of 30 to 60 nm. Suitable flakes were transferred by dry stamping onto silicon substrates with a thermal oxide layer approximately 290 nm thick. This thickness is optimal to maximize the optical contrast of hBN. The surface profile of the flake was measured using PSI to measure the physical thickness of the flake \cite{vogl_atomic_2019}. The Si/SiO$_2$ substrates were equipped with a grid pattern that simplifies the navigation on the chip.

In the next step, SPEs were created through localized electron beam irradiation, a previously developed process that enables the fabrication of high-yield emitter arrays \cite{kumar_localized_2023}. This method has been shown to reliably produce emitter arrays with identical photophysical properties \cite{kumar_localized_2023,kumar2024polarization,doi:10.1021/acsaom.3c00441}: arrays consisting of $10 \times 10$ irradiated spots with a 4$\mu$m pitch were fabricated using a scanning electron microscope (Helios Nanolab G3), as shown in \autoref{OP-PSI-P1}(c). For beam alignment, imaging, and irradiation, the electron beam was accelerated at 3 kV with current of 25 pA. An electron fluence of $1.4\times 10^{13}$ cm$^{-2}$ was used for imaging and localizing the hBN flake, while a higher fluence of $7.7\times 10^{17}$ cm$^{-2}$ was used with a dwell time of 10 s per spot on pre-defined spots.

\subsection{Optical characterization}
The successful creation of quantum emitters was verified using a commercial fluorescence lifetime imaging microscope (PicoQuant MicroTime 200), which also serves as the initial optical characterization prior to any transfer. We have used a 530 nm pulsed laser with repetition rate of 20 MHz and a pulse length of 80 ps. The irritated spots were excited with 50 $\mu$W of average laser power. The photoluminescence (PL) map of the array was produced using a dwell time of 5 ms per pixel. The excitation of the emitters and the collection of PL were performed by a dry immersion objective with 100$\times$ magnification. The numerical aperture (NA) of 0.9 of the objective was selected to enhance the collection of PL. The working distance of the objective was 0.3 mm. As we collected the PL with the same objective, a long pass filter with a cutoff wavelength at 550 nm was used in the collection path to block the excitation laser. The microscopic system is equipped with Hanbury Brown and Twiss (HBT)-type interferometer, comprising two single photon avalanche diode (SPADs, from Micro Photon Devices) in two arms of a 50:50 beam splitter. The photons get detected whether by these two SPADs or a spectrometer (Andor Kymera 328i). With this arrangement we can measure the second-order correlation function and lifetime, as well as the spectrum of the SPEs. A built-in software (SymPhoTime 64) was used to acquire and for fitting (lifetime decay) the data, which takes the instrument response function into account.

\subsection{Pick-up and transfer process}

\begin{figure*} [t]
    \includegraphics[width = 0.9\textwidth]{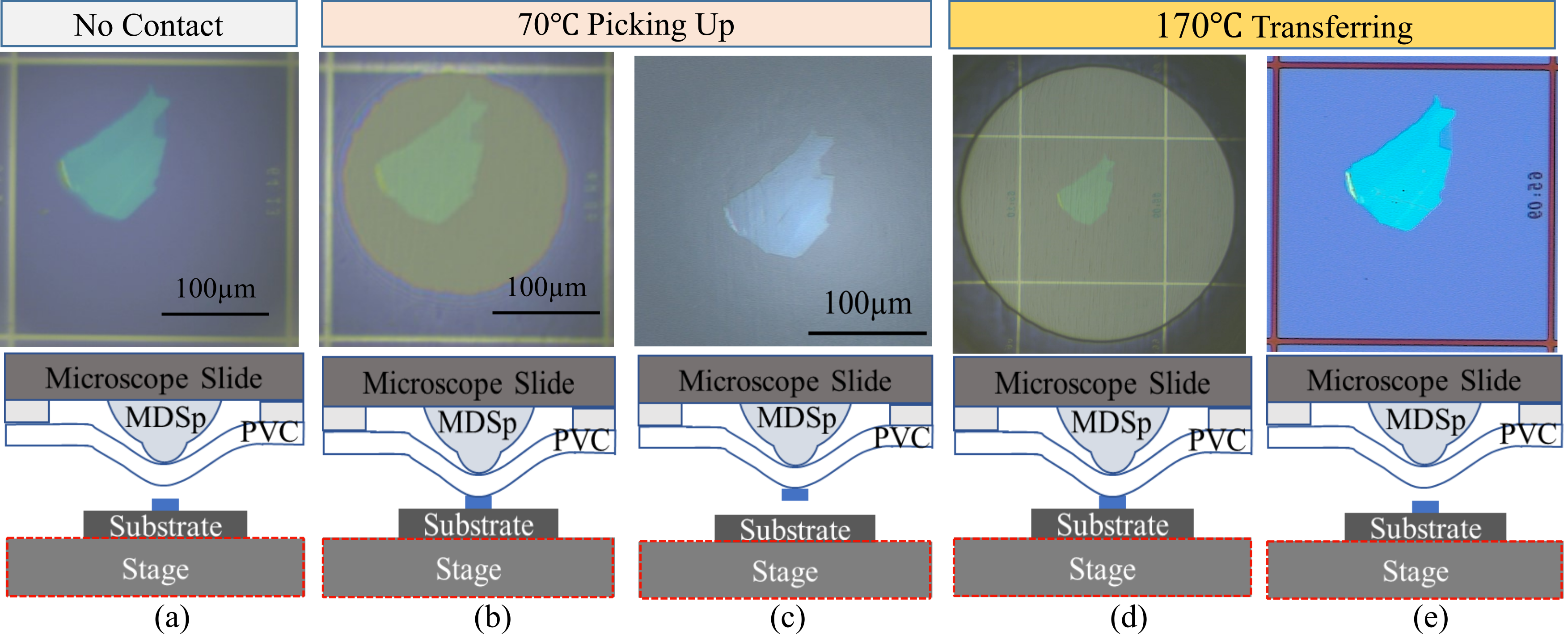}
\captionsetup{
        justification= raggedright, 
        font= small  
    }
    \caption{\textbf{Schematic illustration of PDMS-PVC-based pick-up and subsequent transfer method.} (a) Shows the flake on a substrate before contact with the TTS. (b) At a substrate temperature 70°C, the TTS was brought into contact with the flake. (c) Illustrates that the flake is on the TTS after being peeled off from the substrate. (d) The flake was transferred to different coordinates of the substrate and released at 170°C. (e) Depicts the successfully transferred flake on a different position. The red-dotted box around the stage represents the temperature-controlled hotplate used for heating during the pick-up and transfer process.}
    \label{Figures/Pick up and Place Schematic_Updated.png}
\end{figure*} 

Finally, the emitter arrays were picked up and transferred to new target substrates using TTS to facilitate integration into experimental setups or devices. Wakafuji et al. \cite{wakafuji_3d_2020} developed a method of preparing TTS for precise manipulation of 2D materials for creating van der Waals heterostructures. We have adapted and optimized for our hBN quantum emitter arrays (see Supplementary Section S1 for details). These modifications made the process simple and effective for our experiment. The TTS comprises of two polymer-based components: a base made of PDMS and an adhesion layer of poly(vinyl chloride) (PVC). The base is a microdome-shaped polymer (MDSp), created by sequential stacking of PDMS droplets, designed to achieve a low-curvature structure for effective pickup of thin flakes surrounded by non-essential flakes. The low curvature ensures good contact with the flake being picked up while simultaneously preventing the pickup of nearby flakes. This way we ensure a clean process which does not contaminate other areas of the target chip. The MDSp was then covered with a PVC foil to utilize the temperature-dependent adhesion properties of PVC \cite{wakafuji_3d_2020}.

The PVC film is uses as the adhesion layer which has a temperature-dependent adhesion. It has a glass-transition temperature of around 90°C, a melting point at around 180°C, and it shows formidable adhesion to 2D material flakes already at 70°C \cite{wakafuji_3d_2020}. \autoref{Figures/Pick up and Place Schematic_Updated.png} illustrates the complete pick up and transfer process schematically together with microscope images during all steps: the MDSp domes are mounted on a transparent microscope slide that allow us to image with a CMOS camera and zoom lens from the top to align the TTS above the flake to be transferred (see \autoref{Figures/Pick up and Place Schematic_Updated.png}(a)). The TTS position is controlled by a step motor and is then pressed onto the flake. The contact between them becomes visible through a color change due to the polymer-SiO$_2$ interface, indicating the absence of air in between (see \autoref{Figures/Pick up and Place Schematic_Updated.png}(b)). The sample is then heated to 70 °C with a hot plate from below which makes the hBN to adhere well to the PVC. The TTS is then moved upwards together with the hBN flake (see \autoref{Figures/Pick up and Place Schematic_Updated.png}(c)). After that, a new target substrate could replace the old chip and the TTS is stamped to the new location. For simplicity we have simply moved it to a different location on the same chip with other coordinates in our grid (see \autoref{Figures/Pick up and Place Schematic_Updated.png}(d)). As the substrate is mounted on micromanipulators one can precisely align the flake relative to the chip. In the final step, heating the substrate to 170°C reduces the adhesion between PVC and the hBN flake, allowing the flake to be released while preserving its size and shape (see \autoref{Figures/Pick up and Place Schematic_Updated.png}(e)).

\subsection{Optimization of the transfer parameters}

\begin{table} [t]
\captionsetup{justification=raggedright,singlelinecheck=false} 
\caption{\label{tab:table1} Parameter optimization for hBN flake transfer. The table presents the effect of varying temperature $T$, stage velocity $v$, and contact time $t$ on the success probability of the transfer. Trials are color-coded to indicate unsuccessful (red), low-success (orange), and high-success (green) attempts, highlighting the optimal parameters for reliable flake transfer.}
\begin{ruledtabular}
\resizebox{\columnwidth}{!}{ 
\begin{tabular}{ccccc}
\toprule
Trial & $T$ (°C) &$v$ ($\mu$m/s) & $t$ (min) & Success probability \\
\midrule
\rowcolor{SoftRed} 1  & 100 & 100  & 1 & \textcolor{black}{0} \\
\rowcolor{SoftRed} 2  & 110 & 100  & 1 & \textcolor{black}{0} \\
\rowcolor{SoftRed} 3  & 120 & 10  & 2 & \textcolor{black}{0} \\
\rowcolor{SoftOrange} 4  & 130 & 10  & 2.5 & \textcolor{black}{low} \\
\rowcolor{SoftOrange} 5  & 140 & 10  & 3 & \textcolor{black}{low} \\
\rowcolor{SoftGreen} 6  & 150 & 1 & 3.5 & \textcolor{black}{high} \\
\rowcolor{SoftGreen} 7  & 160 & 1  & 4 & \textcolor{black}{high} \\
\rowcolor{SoftGreen} 8  & 170 & 0.1 & 4.5 & \textcolor{black}{high} \\
\rowcolor{SoftGreen} 9  & 175 & 0.1 & 5 & \textcolor{black}{high} \\
\bottomrule
\end{tabular}
}
\end{ruledtabular}
\end{table}

The flakes were picked up at 70°C in all cases with the fixed velocity of the stage at 10 $\mu$m/s, and the following parameters were iteratively optimized to achieve an optimal transfer outcome:
\begin{itemize}
\item Temperature of the target substrate (from 100°C to 175°C) 
\item Velocity of the transfer stage (from 100 $\mu$m/s down to 0.1 $\mu$m/s) 
\item  Contact time of the TTS and substrate (1 to 5 minutes)
\end{itemize}
We found that increased temperature, extended contact time, and significantly slow velocity yield a significantly higher probability, which is summarized in \autoref{tab:table1}: The success rate of the outcomes of the trials 6-9 (shaded with a green color) was more than 90\% for 8 transfers. The trials 4 and 5 (shaded with an orange color) show a low success probability at moderate temperature, velocity, and slightly higher contact times. Here, the success rate was about 40\% for 10 transfers. On the other hand, the trial 1 to 3 (shaded with a red color) were unsuccessful at lower temperature, shorter dwell time, and higher velocities.

The gradual progression indicates that the correlation between higher temperature, lower velocity of the transfer stage, and the extended dwell time improved the transfer yield. One important point to mention here is that the temperature for each case was increased gradually by 10 degrees, not in a way that the temperature is already set at 170°C and then the transfer was performed.

\begin{figure*} [t]
    \includegraphics[width = 0.9\textwidth]{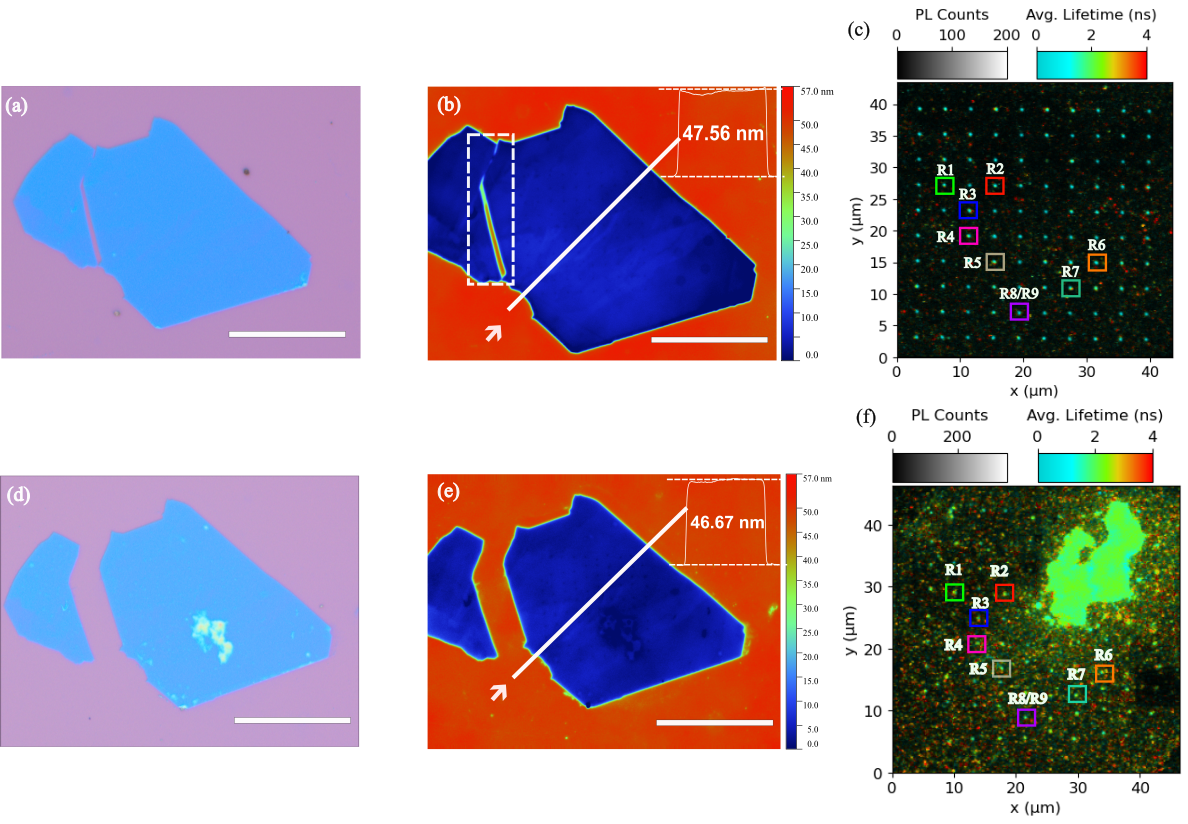 }
\captionsetup{
        justification= raggedright, 
        font= small  
    }
    \caption{\textbf{Overall structural and optical properties of an hBN flake before and after transfer.} (a,d) Show the optical images of the flake before and after the transfer process at 50$\times$ magnification. The scale bar in all images is 20 $\mu$m. Post transfer there becomes some residue visible, as it was inadvertently deposited during the transfer process. The PSI images (b,e) show the surface topography. The thickness of the flake after transfer changed only negligibly (47.56 nm to 46.67 nm), likely caused by the variation of oxide thickness of the substrate. The overall PL maps of the emitter array in (c,f) show the spatial locations of the irradiated spots on the flake. The emitters were excited with a 530 nm laser. The gray scale encodes the fluorescence intensity and the color encodes the average lifetime.}
    \label{OP-PSI-P1}
\end{figure*}

\section{Results} \label{sec:results}

\begin{figure*}[t]
    \includegraphics[width =0.9\textwidth]{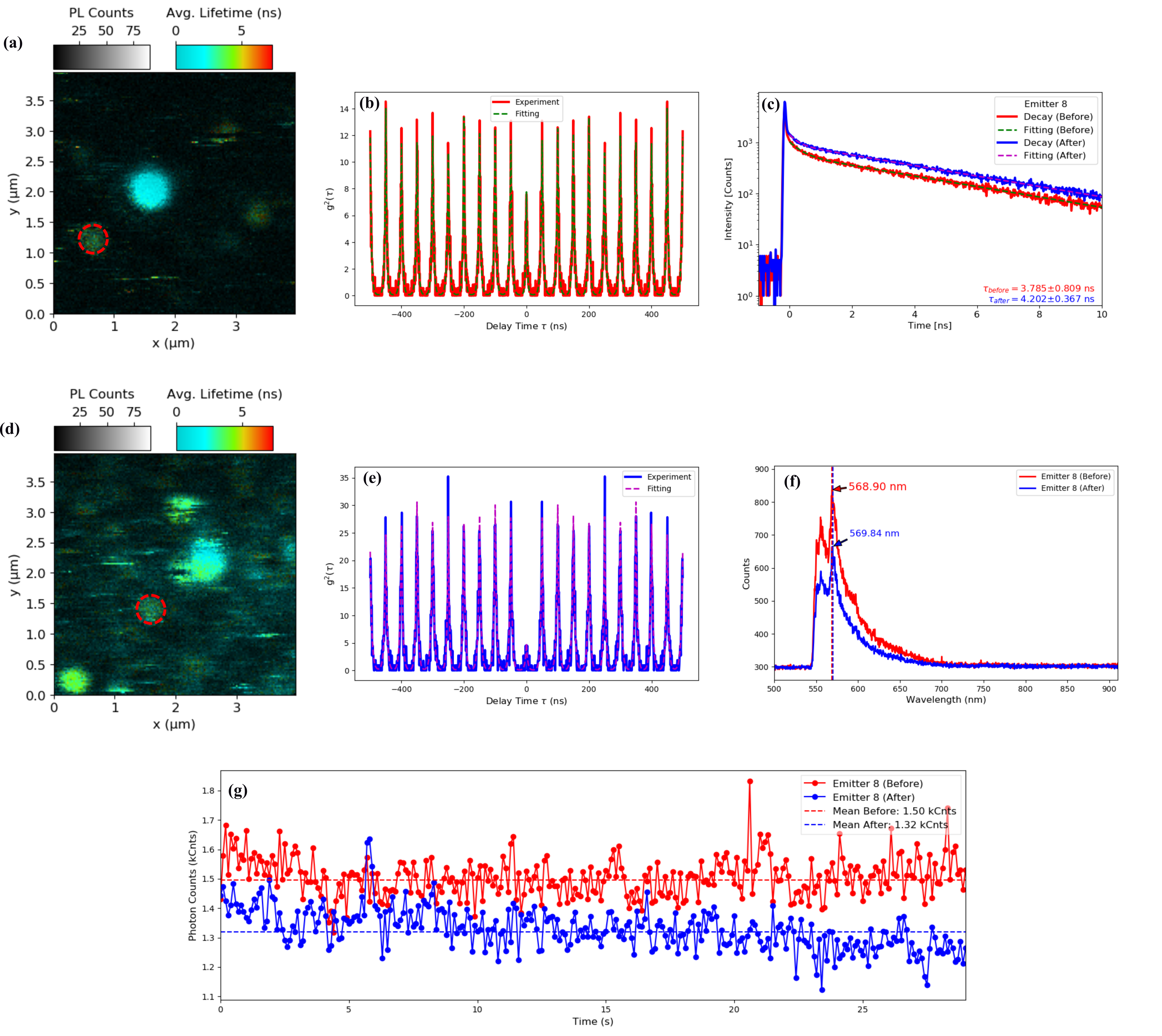 }
    \captionsetup{
        justification=raggedright, 
        font= small  
    }
    \caption{{\textbf{Characterization of Emitter-8 near the irradiated spot R2 before and after transfer.}} (a,d) Show the combined PL and lifetime map before, after the transfer, respectively. The red circle marks the actual emitter from which the following photophysical properties have been measured. (b,e) Show the second-order correlation function $g^{(2)}(0)$ measured with a 20 MHz pulsed laser. The values are 0.674(5) and 0.186(2) for the before and after cases, respectively. The photon purity has therefore significantly improved post-transfer. (c) Shows the lifetime decay curves, with lifetimes of $\tau_{\text{before}} = 3.785(8)$ ns and $\tau_{\text{after}} = 4.202(4)$ ns. A slight change was observed, but both values remained well within the typical variations described in Ref.\ \cite{kumar_localized_2023}. (f) Shows the emission spectrum with a peak at 569.84 and 568.90 nm before and after the transfer, i.e., only a minimal spectral shift. (g) Displays the stability of the photon count over 30 seconds with relative fluctuations before and after transfer being 9.37\% and 9.74\%, respectively. The reduction in photon count after the transfer can be attributed to the polarization and alignment of the laser, as the measurements were taken on different days with slightly different relative alignments.}
    \label{fig:Emitter 8}
\end{figure*}

The above described transfer method can only qualify to be successful if the flakes retain their structural and the emitters their photo-physical properties. Also the former is important, e.g., when integrating the emitters with optical cavities \cite{vogl_compact_2019} where the crystals must not have any cracks that would act as scattering centers out of the cavity mode. We performed a full optical characterization following the fabrication of the device (i.e., before the transfer) and the same characterization was repeated after the pick-up and transfer to evaluate the physical and optical characteristics of the quantum emitters.

We first show the structural and overall optical properties of a transferred hBN flake.
The contrast-enhanced optical microscopic image in \autoref{OP-PSI-P1}(a) shows the hBN flake with a uniform thickness and most of the crystal is free of visible cracks except one major crack outlined with white dotted enclosure in the \autoref{OP-PSI-P1}(b). This is a typical artefact of the mechanical exfoliation process. The surface profile (PSI image) is shown in \autoref{OP-PSI-P1}(b) from which an optical path length (OPL) of 47.56 nm can be extracted. The optical path length is the effective distance that light travels in a given medium where geometric distance and refractive index both are considered. This can be converted to physical thickness via rigorous coupled-wave analysis (RCWA) simulations \cite{vogl_fabrication_2018} and corresponds to a physical thickness of around 30 nm. The lifetime imaging map is shown in \autoref{OP-PSI-P1}(c) where the emitter array is clearly visible. The gray scale of the pixel values encodes the fluorescence intensity, while the color value encodes the lifetime. Several spots are marked with boxes for which detailed photophysical properties are presented later. After picking-up the flake from a specific quadrant in the grid (47:05) and transferring it to a new quadrant (42:05) (see supplementary Section S2), we repeated the same set of measurements. \autoref{OP-PSI-P1}d) shows the post-transfer contrast-enhanced optical image, which retained the shape and size of the flake with no additional visible wrinkles. At the location of the initial crack, however, the flake has been torn apart during the transfer process. When looking at the SEM image with a much higher spatial resolution than the optical images (see Supplementary Section S2), one can actually see that this crack goes through the entire crystal before the transfer. This implies that two separate flakes were transferred which have increased their distance between each other, but individually remained fully intact. The PSI image in \autoref{OP-PSI-P1}(e) indicates no change in surface roughness or thickness which remained at 46.67 nm OPL. It should be noted that this OPL depends on the thickness of the oxide layer below which adds an additional phase shift to the light and there is always a variation of oxide thickness across the entire chip for the thermally grown oxide. Also for all other (successfully classified) transfers we observed that the transfer process preserved the structural integrity of the hBN flakes. For the emitter array shown in \autoref{OP-PSI-P1}(f) we have additionally observed a some fluorescent residue after the transfer. This residue is also visible in the optical microscope image in \autoref{OP-PSI-P1}(d). This could be caused by a dust particle trapped with polymer residue from the PDMS-PVC stamp used during the transfer. This anomaly was observed in only one out of 18 transfers and did not impact the characteristics of the emitters, as confirmed in the remainder of the results section (in Supplementary Section S7 we show an example of a residue-free transfer). Before presenting the details, we would like to clarify the numbering of the emitters before and after transfer. In the pre-transfer characterization, emitters were sequentially numbered starting from 1. However, some emitters were lost during the transfer process as mentioned earlier. To maintain consistency in the comparison before and after transfer, we addressed each emitter with their original number as before.

We now turn to the photophysical properties of the individual emitters before and after the transfer. We showcase two exemplary emitters (more data can be found in the Supplementary Sections S3-S6), localized near the spots R2 and R3 in the fluorescence map in \autoref{OP-PSI-P1}(c). Note that we distinguish between spots labeled R$n$ and Emitter-$m$ (where $n,m$ are consecutive numbers) as in the vicinity of each irradiated spot sometimes can be more than one emitter \cite{kumar_localized_2023}. Each spot is individually characterized in terms of their second-order correlation function $g^{(2)}(\tau)$, excited-state lifetime, spectrum, and photostability. The before measurements are always plotted in red, while the data plotted in blue belong to the measurements after the transfer.

The first emitter presented in detail, Emitter-8, is shown in \autoref{fig:Emitter 8}. The fluorescence lifetime map before (see \autoref{fig:Emitter 8}(a)) shows a nicely localized emitter near the irradiated spot. Prior to the transfer, this emitter exhibited a $g^{(2)}(0)$ value of 0.674(2) (see \autoref{fig:Emitter 8}(b)), and an excited-state lifetime of 3.785(8) ns (see \autoref{fig:Emitter 8}(c)). The fluorescence lifetime map after (see \autoref{fig:Emitter 8}(d)) the transfer shows some randomly distributed additional background. This is likely being caused by the contamination mentioned above, as R2 is very close to the fluorescent residue. The emitter itself, however, maintained its brightness and had a slightly longer lifetime with 4.202(4) ns, as is evident from \autoref{fig:Emitter 8}(c) and (d). The $g^{(2)}(0)$ value significantly improved to 0.186(2). We will discuss potential explanations below when considering the statistics of many emitters. The emission spectra showed negligible change, with peak wavelengths shifting less than 2 nm (see \autoref{fig:Emitter 8}(f)). Also the photostability was maintained, with intensity fluctuations (standard deviation divided by mean, for details about this calculation see Supplementary Section S9) did not change significantly from 9.37\% before to 9.74\% after the transfer (see \autoref{fig:Emitter 8}(g)).

\begin{figure*} [t]
    \includegraphics[width =0.9\textwidth]{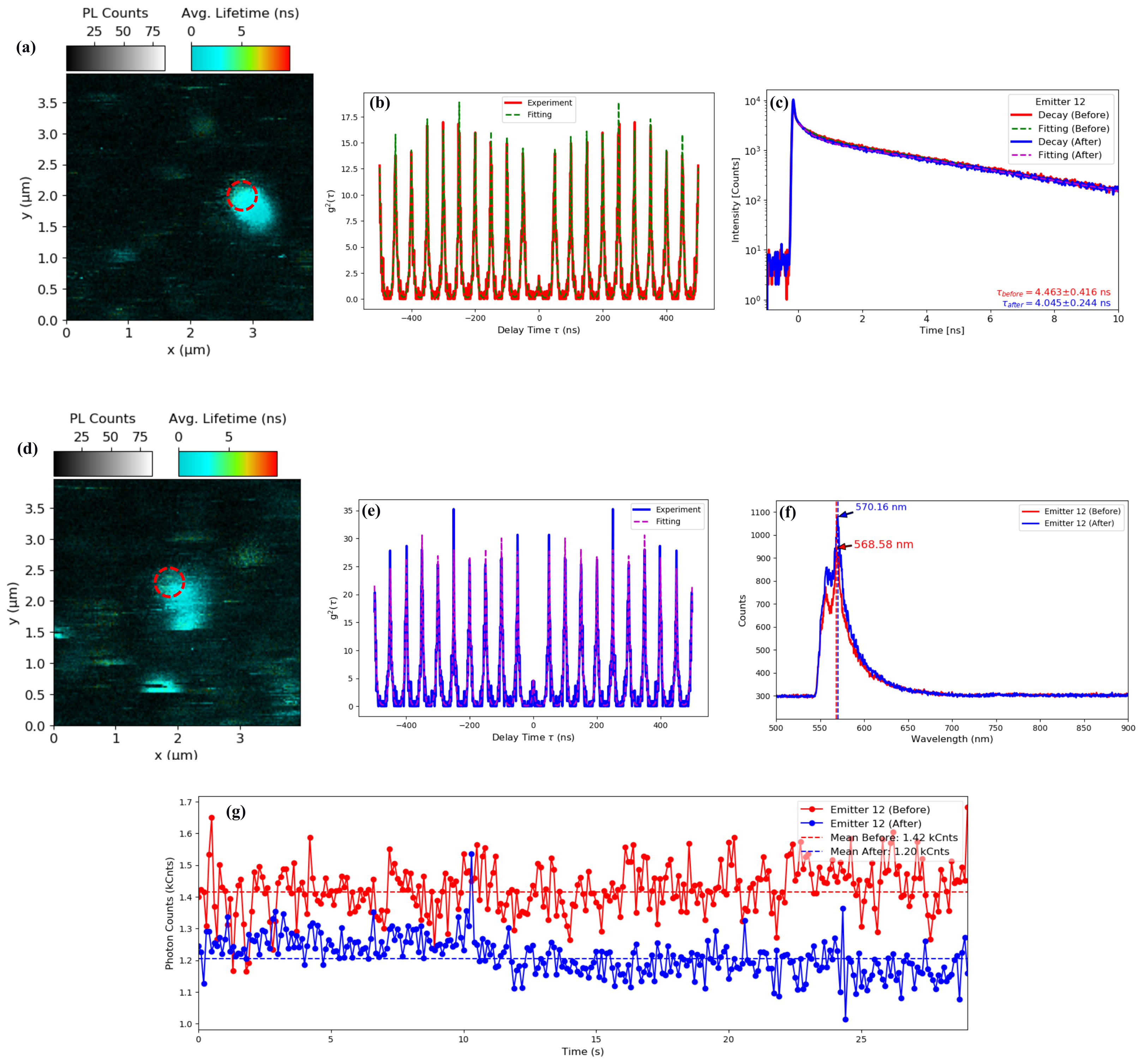}
    \captionsetup{
        justification=raggedright, 
        font= small  
    }
    \caption{{\textbf{Characterization of Emitter-12 near the irradiated spot R3 before and after transfer.}} (a,d) Show the combined PL and lifetime map before, after the transfer, respectively. The red circle marks the actual emitter from which the following photophysical properties have been measured. (b,e) Show the second-order correlation function $g^{(2)}(0)$ measured with a 20 MHz pulsed laser. The values are 0.115(1) and 0.106(5) for the before and after cases, respectively. (c) Shows the lifetime decay curves, with lifetimes of $\tau_{\text{before}} = 4.46(4)$ ns and $\tau_{\text{after}} = 4.045(2)$ ns. (f) Shows the emission spectrum with a peak at 570.16 and 568.58 nm before and after the transfer, i.e., only a minimal spectral shift. (g) Displays the stability of the photon count over 30 seconds with relative fluctuations before and after transfer being 9.82\% and 9.57\%, respectively. The reduction in photon count after the transfer can be attributed to the polarization and alignment of the laser, as the measurements were taken on different days with slightly different relative alignments.}    
    \label{fig:Emitter 12_Large}
\end{figure*}

A very similar trend can be observed for Emitter-12, as depicted in \autoref{fig:Emitter 12_Large}(a-g): while lifetime, spectrum, and photostability only exhibit a small or negligible change before and after the transfer, the $g^{(2)}(0)$ value of 0.115(1) from before (see \autoref{fig:Emitter 12_Large}(b)) decreased to 0.106(5) after the transfer (see \autoref{fig:Emitter 12_Large}(e)), again indicating an increased photon purity. The lifetime remained steady at around 4.46(4) ns before and 4.045(2) ns after the transfer (see \autoref{fig:Emitter 12_Large}(c)). The emission spectrum showed minimal shift (see \autoref{fig:Emitter 12_Large}(f)), and the stability was again maintained with intensity fluctuations of 9.82\% before and 9.57\% after transfer (see \autoref{fig:Emitter 12_Large}(g)).

\begin{figure*}[t]
     \includegraphics[width =1\textwidth]{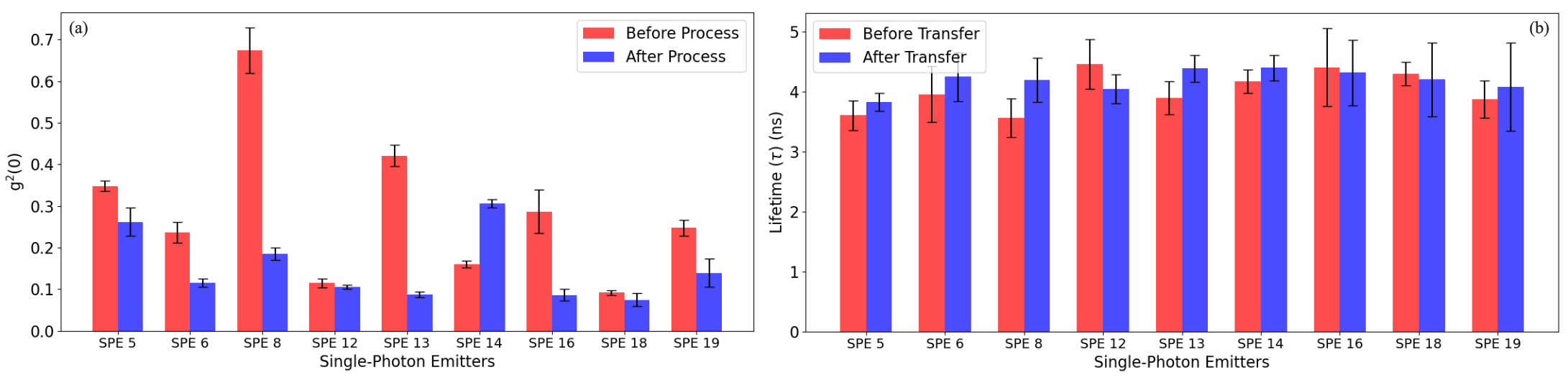 }
    \captionsetup{
        justification=raggedright, 
        font= small  
    }
    \caption{\textbf{Statistics and comparison of properties before and after the transfer process.} Red bars show the values before and blue bars after the transfer.(a) Distribution of $g^{(2)}(0)$ values for the transferred emitter array presented in the main text. The after values are except SPE 14 better than the before values. (b) Distribution of lifetimes for the transferred array presented in the main text. The overall averages remain roughly constant with 4.03(1) and 4.19(1) ns before and after, respectively. The error bars indicate the standard deviation from fitting for each emitter.}
    \label{fig:G2_Lifetime_Comparison}
\end{figure*}

An increase of photon purity is unexpected. The question is, if this is a statistically significant effect. It is known that the $g^{(2)}(0)$ value is influenced by the excitation laser polarization \cite{kumar2024polarization}. Essentially, if the laser polarization is aligned with the excitation dipole, the emitter brightness is maximized while simultaneously all cross-polarized noise sources are suppressed. In \autoref{fig:G2_Lifetime_Comparison}(a) we show the $g^{(2)}(0)$ values of several emitters before and after the transfer with a clear trend toward better $g^{(2)}(0)$ values after the transfer. The average decrease $g^{(2)}(0)$ values (not counting Emitter-14 which was the only emitter for which there was an increase) post-transfer is 45.92\%. Therefore it is unlikely that this is purely caused by a better polarization alignment (e.g., because the sample is rotated relatively to the laser polarization after the transfer). If this was the case, we would expect that for some emitters this alignment improves, while for other it decreases, because the dipole orientation can only take certain values relative to the crystal axes \cite{kumar2024polarization}. As we have used a temperature of 170°C during the transfer process, it could also be possible that some contaminants (noise sources) burned off and as a result, the photon purity increased. Another reason might be that due to the transfer process, there is a change in stress and strain within the flake which can change the transition rates in multi-level systems and therefore also change the $g^{(2)}(0)$ value. Without further experiments, however, these potential explanations will remain speculative and should be explored in future work. On the other hand, the lifetime values remain much consistent after transfer (see statistics in \autoref{fig:G2_Lifetime_Comparison}(b)): The average lifetime across all emitters in that array remained consistent with average value of $4.03(1)$ ns and $4.19(1)$ ns, respectively, before and after transfer. This indicates that there is no significant change in radiative or non-radiative decay channels.

In the assessment of the efficiency of our method, we did not consider the emitters which were lost due to the coincidence of fluorescent residue deposition on the hBN flake. Before the transfer, 20 emitters on this flake were initially measured, and 13 remained measurable after transfer (i.e., 7 of them were lost due to the residue). Among these 13 emitters, 11 showed typical photophysical properties before the transfer (the two unusual ones were not further considered). After the transfer, we successfully measured those 11 emitters, with 9 retaining consistent photophysical properties, including the second-order correlation function, lifetime, spectrum, and photostability. This results in a success probability of 81.8\%. As already mentioned, in this main text we only present the results of one emitter transfer. Further data of individual emitters on that array and other successful transfers can be found in the Supplementary Sections S3-S9.

\section{Conclusions}\label{sec:conclusions}
We have developed an all-dry transfer technique of quantum emitters hosted by hBN crystals. Our transfer technique stands out for its remarkable simplicity and efficiency. It operates without involving any chemical solvents during the transfer process, ensuring a clean method that requires no post-cleaning. A significant advantage is the durability of the PDMS-PVC stamp; unless it is melted or otherwise damaged, a single stamp can be re-used for several transfers, with only the adhesion layer needing replacement which is a straightforward and uncomplicated task. The tunable curvature area of the PDMS droplet adds versatility, making it highly convenient for different device configurations. Moreover, the process works with a relatively low temperature of 170°C, which is easily achievable with standard hot plates. In addition this is way below the activation temperature at which defects in hBN become mobile \cite{vogl_atomic_2019}. Collectively, these features make our transfer technique practical, adaptable, and efficient for a wide range of applications. This is in particular useful when the emitter fabrication method requires special substrates, such as conducting substrates for localized charged particle irradiation. Our method has successfully demonstrated on Si/SiO$_2$ substrates. As Si and silicon-on-insulator (SOI) structures are widely used for fabricating waveguides and other photonic devices \cite{rahim2017expand, YoungbloodLi+2017+1205+1218}, we expect that our method can be broadly used in these cases. Furthermore, 2D materials adhere strongly to gold surfaces \cite{torres2018adhesion}, whereas PVC's adhesion becomes weaker at higher temperature. Therefore, our transfer method is expected to work well on gold-coated substrates. We also expect that the method can be easily adapted for quantum emitters in other 2D materials such as transition metal dichalcogenides \cite{Tonndorf:15,nnano.2015.60,nnano.2015.67,he2015single,nnano.2015.79}. Therefore, our work provides an important step for the advancement of applying single photon emitting defects in quantum technology applications.


\section*{Acknowledgements} \label{sec:acknowledgements}

This research is part of the Munich Quantum Valley, which is supported by the Bavarian state government with funds from the Hightech Agenda Bayern Plus. This work was funded by the Deutsche Forschungsgemeinschaft (DFG, German Research Foundation) - Projektnummer 445275953, under Germany's Excellence Strategy - EXC-2111-390814868, and as part of the CRC 1375 NOA project C2.


\section*{Notes}
The authors declare no competing financial interest.

 


\bibliography{main}
\end{document}